\documentstyle[11pt]{article}

\begin{document}

\title{Quantum integrability of sigma models\\
on $AII$ and $CII$ symmetric spaces.}
\author{\textbf{A.Babichenko\thanks{%
address:Shahal str. 73/4, Jerusalem, 93721, Israel;
e-mail:ababichenko@hotmail.com}}}
\date{October 2002}
\maketitle

\begin{abstract}
Exact massive S-matrices for two dimensional sigma models on symmetric
spaces $SU(2N)/Sp(N)$ and $Sp(2P)/Sp(P)\times Sp(P)$ are conjectured. They
are checked by comparison of perturbative and non perturbative TBA
calculations of free energy in a strong external field. We find the mass
spectrum of the models and calculate their exact mass gap.
\end{abstract}

\section{Introduction}

   Recently renewed interest to two dimensional sigma models
with and without topological theta terms on different coset and, in
particular, symmetric and supersymmetric spaces was stimulated by their
relevance for Integer Quantum Hall effect \cite{Zirn} - \cite{RL}, and two
dimensional fermionic systems with different types of symmetry and quenched
 disorder (see, for example,\cite{GW} - \cite{GLL},\cite{Zirn1}). This is
the reason why any new exact solution of such models, being important by
itself in general, can also find a practical application.

Essential part of symmetric space sigma models are exactly solvable, since
were proved (or believed) to be integrable. Classical integrability of sigma
models on any symmetric spaces $G/H$ is known for a long time \cite{FK}, but
quantum integrability for symmetric space with $H$ non simple may be
destroyed by possible anomalies \cite{AGF}. But even for $H$ simple, when
quantum integrability is undoubted, exact S-matrices are known not for all
symmetric spaces. (For Cartan classification of symmetric spaces see for
example \cite{Helg}).

Recently \cite{FenTeta} the list of known exact S-matrices for integrable
sigma models on symmetric spaces was extended by new massive and massless
S-matrices for spaces $AI$ ( $SU(N)/SO(N)$ ) and $BDI$ ( $O(2P)/O(P)\times
O(P)$ ) without and with theta term. The conjectured S-matrices were checked
by comparison of $T=0$ TBA calculations of free energy of the system in
strong external field with perturbative calculations \cite{FenTeta}, and
also by $T\rightarrow \infty $ TBA extraction of UV central charge for these
two models \cite{FenCoset}. It is remarkable that $BDI$ sigma model turns
out to be quantum integrable in spite of possibility of anomalies.

There are other close relatives of the models considered by Fendley: sigma
model on $AII$ ( $SU(2N)/Sp(N)$ ) is close to $AI$, and sigma model on $CII$
( $Sp(2P)/Sp(P)\times Sp(P)$ ) is analogous to $BDI$ sigma model - in both
cases orthogonal (sub)group is changed to symplectic. We are going to show
that the analogy extends also to S-matrix conjecture, providing one more
example ($CII$) of quantum integrability of non supersymmetric sigma model
on a factor group manifold with non simple invariant subgroup.

The plan of the discussion is the following. The first section describes
standard perturbation theory technique of free energy calculation in a
strong external field. In the second section we conjecture our fundamental
S-matrices for the models using some symmetry arguments in support of it,
and show how the mass spectrum of the model follows from bootstrap. After
that we calculate the same free energy using $T\rightarrow 0$ TBA technique
based on the conjectured exact S-matrix. We show correspondence between
perturbative and non perturbative results which confirms correctness of the
conjectured S-matrices. Moreover, two expressions for the same quantity -
the free energy - fixes the mass gap for the models exactly. We conclude by
brief discussion of results and give some technical details for perturbative
calculations and the S-matrices, like explicit form of projectors, in the
Appendix.

\section{Perturbative analysis}

\bigskip We are considering the Lagrangian 
\[
{\cal L}_{0}=\frac{1}{\lambda ^{2}}Tr\{\partial _{\mu }\Phi \partial
_{\mu }\Phi ^{\dagger }\} 
\]
where we introduced a matrix representation $\Phi $ for coset $G/H$
elements. Important property of this Lagrangian is its global $G$
invariance. We work with Lie algebraic current representation of Lagrangian: 
\begin{equation}
{\cal L}_{0}=-\frac{1}{\lambda ^{2}}Tr\{(g^{-1}\partial _{\mu
}g)(g^{-1}\partial _{\mu }g)\}  \label{Lagr}
\end{equation}
and consider representation of a coset group element as an
exponentialization of its Lie algebra: $g=\exp
(i\sum_{I}n_{I}E^{I}+i\sum_{i}n_{i}H_{i})$, where $H_{i}$ are generators of
Cartan subalgebra of a coset space, and $E^{I}$ - other its generators.
Explicit form of the basis $E_{I}$ we use for calculations one can find in
the Appendix.

In order to check our conjectures about an S-matrix of the model (see the
next section), we do, for today, a standard procedure of putting the system
into a strong external field \cite{Pert}. Strong field (and hence high
energy limit) gives an opportunity to believe to perturbation theory,
because of asymptotical freedom of the models. So we replace derivatives in
(\ref{Lagr}) by ''covariant'' ones: 
\[
D_{\mu }g=\partial _{\mu }g-h\delta _{\mu 0}(Qg+gQ)
\]
where $h$ is a strength of the external field, which is chosen in the
direction $\overrightarrow{q}$ in the Cartan subalgebra of $G$: $%
Q=\sum_{i=1}^{r}q_{i}H_{i}=(qH)$. The lagrangian density (\ref{Lagr}) in the
presence of the source becomes 
\begin{equation}
{\cal L}={\cal L}_{0}+\frac{2h}{\lambda ^{2}}Tr\{\left( g^{-1}Q+Qg^{-1}\right)
\partial _{0}g\}-\frac{2h^{2}}{\lambda ^{2}}Tr\{Q^{2}+g^{-1}QgQ\}
\label{Lag2}
\end{equation}
We are going to calculate the dependence of the free energy on the external
field $h$: $\delta f(h)=f(h)-f(0)$ using perturbative calculation in the
running coupling constant $\lambda (h)$. We will restrict ourselves by
quadratic part of the euclidian lagrangian in the fields $n_{I}$, which
turns out to be enough for our purposes. Some details of calculations one
can find in Appendix and the result is 
\[
{\cal L}\simeq -\frac{4h^{2}}{\lambda ^{2}}q^{2}+\frac{1}{\lambda ^{2}}%
{\cal L}_{0}^{\prime }
\]
where for both $AII$ and $CII$ cases 
\begin{equation}
{\cal L}_{0}^{\prime }=\frac{1}{2}\sum_{i\geq j=1}^{N}\{n_{ij}((\partial_\mu)^2
+h^{2}M_{ij}^{I}(q))n_{ij}^{\ast }+m_{ij}((\partial_\mu)^2
+h^{2}M_{ij}^{II}(q))m_{ij}^{\ast }\}  \label{Lag0}
\end{equation}
with mass matrices 
\begin{eqnarray*}
M_{ij}^{I}(q) &=&(q_{i}-q_{j})^{2}+(q_{i+N}-q_{j+N})^{2} \\
M_{ij}^{II}(q) &=&(q_{i}-q_{j+N})^{2}+(q_{i+N}-q_{j})^{2}
\end{eqnarray*}
for $AII$ case and 
\[
M_{ij}^{I}(q)=(q_{i}-q_{j+P})^{2},\ M_{ij}^{II}(q)=(q_{i}+q_{j+P})^{2}
\]
for $CII$ case. (Here in (\ref{Lag0}) there is no need in complex
conjugation * in $CII$ case.)

Lets point out that at this level some of the fields $n_{I}$ decoupled, and
we wrote only those which enter in an $h$ dependent manner. At the tree level
one has $\delta f(h)_{0}=-\frac{4h^{2}}{\lambda ^{2}}q^{2}$, $%
q^{2}=\sum_{i=1}^{2N(2P)}q_{i}^{2}$. Free energy at the one loop level is
just properly regularized $\frac{1}{2}\sum_{i,j}\ln \det ((\partial_\mu)^2
+h^{2}M_{ij}(q))$.Using standard dimensional regularization $\varepsilon =d-2
$ one finds 
\[
\delta f(h)_{1}=-\frac{h^{2}\beta _{1}}{2\pi \varepsilon }q^{2}+\frac{h^{2}}{%
4\pi }\sum_{I}\sum_{i\geq j=1}M_{ij}^{I}(q)\left[ 1-\gamma _{E}+\ln 4\pi
-\ln \left( \frac{h^{2}}{\mu ^{2}}M_{ij}^{I}(q)\right) \right] 
\]
where $\mu $ is a mass parameter of dimensional regularization, and $\beta
_{1}q^{2}=\sum_{I}\sum_{i\geq j=1}M_{ij}^{I}(q)$. After some algebra one can
find that $\beta _{1}=2N$ for $AII$ case, and $\beta _{1}=2P+1$ for $CII$
case. Here we use the condition $\sum_{1}^{2N(2P)}q_{i}=0$, which is
necessary in $AII$ case, and just will correspond to our concrete choice of
the external field in $CII$ case (see below).

The point is that the quantity $\delta f(h)$ is renormalization group
invariant when $\lambda $ runs with $\mu $, so we can set $\mu =h$ and use
the results of $\beta $-function calculations (without external field), done
for almost all symmetric spaces \cite{Hik} up to three loops. We need the
result up to two loops: 
\begin{equation}
h\frac{\partial }{\partial h}(\lambda ^{2})=-\frac{1}{8\pi }(\beta
_{1}\lambda ^{4}+\beta _{2}\lambda ^{6})-O(\lambda ^{8})  \label{RGE}
\end{equation}
where $\beta _{2}=2N(N-1)$ for $AII$ and $\beta _{2}=P(2P+1)$ for $CII$, and 
$\beta _{1}$ is the same as above, since our calculation reproduced the
correct form of one loop beta function. So adding necessary counterterm to
lagrangian we get the following expression for the free energy 
\begin{equation}
\delta f(h)=-\frac{4h^{2}}{\lambda (h)^{2}}q^{2}-\frac{h^{2}}{4\pi }%
\sum_{I}\sum_{i\geq j=1}M_{ij}^{I}(q)[\ln M_{ij}^{I}(q)-1]+O(\lambda ^{2})
\label{Freen}
\end{equation}
\bigskip One can solve equation (\ref{RGE}) 
\[
\frac{1}{\lambda ^{2}(h)}=\beta _{1}\ln \frac{h}{\Lambda _{MS}}+\frac{\beta
_{2}}{\beta _{1}}\ln \ln \frac{h}{\Lambda _{MS}}+O\left( \frac{1}{\ln \frac{h%
}{\Lambda _{MS}}}\right) 
\]
where $\Lambda _{MS}$ is the cutoff parameter of minimal subtraction scheme,
and substitute it into (\ref{Freen}): 
\begin{equation}
\delta f(h)=-4h^{2}q^{2}\beta _{1}\left( \ln \frac{h}{\Lambda _{MS}}+\frac{%
\beta _{2}}{\beta _{1}^{2}}\ln \ln \frac{h}{\Lambda _{MS}}+c\right) 
\label{FreenL}
\end{equation}
\begin{equation}
c=\sum_{J}\sum_{i\geq j=1}M_{ij}^{J}(q)[\ln M_{ij}^{J}(q)-1]  \label{C}
\end{equation}
This expression will be used for comparison with result of free energy
calculation by TBA based on exact S-matrix, which we are going to present
now.

\section{Exact S-matrices}

As in many other examples of quantum integrable models with higher rank Lie
algebraic (actually Yangian) symmetries, one can expect that particles group
into multiplets corresponding to irreducible representations of the
symmetry. As we mentioned, there is a global $G$ symmetry acting on the
coset space $G/H$, so we assume the S-matrix is related to branching rules
of decomposition of highest weight reps of $G$ into irreducible reps of $H$.
The fundamental S-matrix usually is related to the shortest highest weight
reps of $G$. In \cite{Zirn} a general matrix form was suggested, to which
one can transform any factor group element of symmetric spaces by choice of 
a proper $H$ gauge. These forms are quite useless for us here, but one can see 
that these matrix forms may be done antisymmetric for both $AII$ and $CII$
cases. The minimal antisymmetric representations in $su$ and $sp$ algebras
are reps with highest fundamental weight $\mu _{2}$.
This gives rise to our conjecture: the
fundamental S-matrix of $AII$ and $CII$ symmetric space sigma models are
described by rational $\mu _{2}\times \mu _{2}$ S-matrices of $SU(2N)$ and $%
Sp(2P)$ symmetry correspondingly. As it is well known, Lie algebraic
symmetry with crossing \ and unitarity does not fixes S-matrix completely.
The remaining so called CDD ambiguity is very important, since it in
particular may change the pole structure of the S-matrix, i.e. defines bound
states and spectrum of the model. This CDD ambiguity should be resolved
using any kind of arguments, e.g. physically required coincidence of the
S-matrix to a some known one, at a specific value of one of its parameters.
There are two types of rational S-matrices of general series of Lie
algebraic symmetries. Gross-Neveu like S-matrices have additional CDD
factors with poles which, through the bootstrap, lead to a set of massive
multiplets corresponding to all highest fundamental weights reps, while
sigma model like S-matrices usually are ''minimal'' (have no poles in the
physical strip of rapidity) and have no these CDD factors. As it was
conjectured and confirmed by different checks \cite{FenTeta}\cite{FenCoset},
sigma models on the symmetric spaces $AI$ and $BDI$, are similar rather to
Gross Neveu models, since they have bound state coming from CDD poles. The
same happens in our case.

The fundamental S-matrix of $AII$ sigma model has the form 
\begin{equation}
S_{\mu _{2},\mu _{2}}(\theta )=X(x)S_{\min }(x)\left( {\cal P}_{2\mu
_{2}}+\frac{\Delta +x}{\Delta -x}{\cal P}_{\mu _{3}+\mu _{1}}+\frac{%
\Delta +x}{\Delta -x}\frac{2\Delta +x}{2\Delta -x}{\cal P}_{\mu
_{4}}\right)  \label{Susm}
\end{equation}
with 
\begin{eqnarray}
X(x) &=&-\frac{\sinh \frac{1}{2}(\theta +4\pi i/2N)}{\sinh \frac{1}{2}%
(\theta -4\pi i/2N)}=\frac{\sin \pi (2\Delta +x)}{\sin \pi (2\Delta -x)}, 
\nonumber \\
S_{\min }(x) &=&\frac{\Gamma (1-x)}{\Gamma (1+x)}\frac{\Gamma (\Delta +x)}{%
\Gamma (\Delta -x)}\frac{\Gamma (1-\Delta -x)}{\Gamma (1-\Delta +x)}\frac{%
\Gamma (2\Delta +x)}{\Gamma (2\Delta -x)}
\end{eqnarray}
where $x=\frac{\theta }{2\pi i}$, $\Delta =\frac{1}{2N}$, and ${\cal P}%
_{\omega }$-projector on a rep with highest weight $\omega $. Explicit form
of the projectors one can find in Appendix.

For the $CII$ sigma model we conjecture the following form of the
fundamental S-matrix

\begin{eqnarray}
S_{\mu _{2},\mu _{2}}(x) &=&X(x)S_{\min }(x)\left( {\cal P}_{2\mu _{2}}+%
\frac{\frac{1}{2}-\Delta +x}{\frac{1}{2}-\Delta -x}{\cal P}_{2\mu _{1}}+%
\frac{\frac{1}{2}-\Delta +x}{\frac{1}{2}-\Delta -x}\frac{\frac{1}{2}+x}{%
\frac{1}{2}-x}{\cal P}_{0}\right.  \label{Spsm} \\
&&\left. +\frac{\Delta +x}{\Delta -x}{\cal P}_{\mu _{1}+\mu _{3}}+\frac{%
\Delta +x}{\Delta -x}\frac{\frac{1}{2}-\Delta +x}{\frac{1}{2}-\Delta -x}%
{\cal P}_{\mu _{2}}+\frac{\Delta +x}{\Delta -x}\frac{2\Delta +x}{2\Delta
-x}{\cal P}_{\mu _{4}}\right)  \nonumber \\
S_{\min }(x) &=&\frac{\Delta +x}{\Delta -x}\frac{\Gamma (-\Delta -x)}{\Gamma
(-\Delta +x)}\frac{\Gamma (2\Delta +x)}{\Gamma (2\Delta -x)}\frac{\Gamma
(1-x)}{\Gamma (1+x)}\frac{\Gamma (\Delta +x)}{\Gamma (\Delta -x)}\times  \nonumber
\\
&&\times\frac{\Gamma (1+x)}{\Gamma (1-x)}\frac{\Gamma (\frac{1}{2}+\Delta -x)}{%
\Gamma (\frac{1}{2}+\Delta +x)}\frac{\Gamma (\frac{1}{2}-\Delta +x)}{\Gamma (%
\frac{1}{2}-\Delta -x)}\frac{\Gamma (\frac{1}{2}+2\Delta -x)}{\Gamma (\frac{1%
}{2}+2\Delta +x)}  \nonumber \\
X(x) &=&\frac{\sin \pi (x+2\Delta )}{\sin \pi (x-2\Delta )}\frac{\sin \pi
\left( x+\frac{1}{2}-2\Delta \right) }{\sin \pi \left( x-\frac{1}{2}+2\Delta
\right) }  \nonumber
\end{eqnarray}
where $\Delta =\frac{1}{2(2P+1)}$.

In both cases \bigskip the product $S_{\min }$ and parenthesis is the
minimal unitary and crossing symmetric S-matrices of $SU(2N)$ and $Sp(2P)$
symmetry, fused from corresponding elementary vector representation
S-matrices \cite{ORW}\cite{KRS}\cite{HollS}. They have no poles on the
physical strip of rapidities $\theta $. Additional CDD factors in both cases
provide the only pole (and hence a bound state particle) in the ${\cal P}%
_{\mu _{4}}$ channels (in $Sp$ case there is also cross channel pole). It is
clear that S-matrix describing the scattering of particle from fundamental $%
\mu _{2}$ multiplet on the particle from $\mu _{4}$ multiplet may be
obtained by fusion and will give a pole in $\mu _{6}$ projector, and so on.
In this way we get a spectrum for both models described by $\mu _{2k}$
multiplets,( $k=1,...,N-1$ for $AII$ case and $k=1,...,P$ for $CII$ case).
Mass spectrum can be written as $m_{k}=M\sin \left( \pi \frac{k}{N}\right) $
for $AII$, and $m_{k}=M\sin \left( \pi \frac{2k}{2P+1}\right) $ for $CII$,
where $M$ is a mass scale.

As it sometimes happens in series of higher rank symmetric integrable
models, in their lowest rank cases they often coincide with some other
series (for example, lowest rank thermally perturbed $WD_{n}$ CFT $n=2$ are
just parafermions with a proper perturbation). A remarkable hint for
integrability of the models we are considering here, we get from the fact
that $SU(4)/Sp(2)$ is isomorphic to $SO(6)/SO(5)$. It means that at $N=2$
our $AII$ S-matrix should have the well known form of $O(6)$ sigma model. In
the same way for $CII$ case $Sp(2)/Sp(1)\times Sp(1)\sim
SO(5)/SU(2)\times SU(2)\sim SO(5)/SO(4)$ - it is the $O(5)$ sigma model.
Lets recall that $O(K)$ sigma model has only one (vector) multiplet of
particles in the spectrum (no bound states) with the following S-matrix 
\begin{eqnarray*}
S^{O(K)}(\theta ) &=&\frac{\Gamma (1-x)\Gamma (\frac{1}{2}+x)}{\Gamma
(1+x)\Gamma (\frac{1}{2}-x)}\frac{\Gamma (x+\frac{1}{K-2})\Gamma (\frac{1}{2}%
+\frac{1}{K-2}-x)}{\Gamma (-x+\frac{1}{K-2})\Gamma (\frac{1}{2}+\frac{1}{K-2}%
+x)}\left( {\cal P}_{S}+\frac{x+\frac{1}{K-2}}{x-\frac{1}{K-2}}{\cal P}%
_{A}\right. \\
&&\left. +\frac{x+\frac{1}{K-2}}{x-\frac{1}{K-2}}\frac{x+\frac{1}{2}}{x-%
\frac{1}{2}}{\cal P}_{0}\right)
\end{eqnarray*}
After some $\Gamma $ function algebra one can see that $K=6$ case really
coincides with (\ref{Susm}) at $N=2$ , with corresponding representations
mapping ${\cal P}_{S}\rightarrow {\cal P}_{2\mu _{2}},{\cal P}%
_{A}\rightarrow {\cal P}_{\mu _{1}+\mu _{3}},{\cal P}_{0}\rightarrow 
{\cal P}_{\mu _{4}}$. In the same way one can check that $K=5$ S-matrix
is the same as (\ref{Spsm}) at $P=1$. In this case representation
correspondence has the form ${\cal P}_{S}\rightarrow {\cal P}_{2\mu
_{2}},{\cal P}_{A}\rightarrow {\cal P}_{2\mu _{1}},{\cal P}%
_{0}\rightarrow {\cal P}_{0}$. (Projectors ${\cal P}_{\mu _{1}+\mu
_{3}},{\cal P}_{\mu _{2}},{\cal P}_{\mu _{4}}$ are absent in this
case.).

\section{\protect\bigskip TBA calculation of the free energy}

The main point in $T=0$ TBA analysis of our models in external field is
based on a skill to chose external field in such a way, that TBA system will
be the simplest, i.e. the ground state will contain the minimal number of
particles generated from vacuum by external field. The fact that the field
is strong gives a basis for the assumption that only particles with maximal
charge will be generated by external field. As we said the fundamental
S-matrices for both $AII$ and $CII$ models are rank two antisymmetric
tensors $a_{ij}$ and an external field from the Cartan subalgebra of $G$ ,$%
A=diag_{2N,2N}\{A_{1},...,A_{2N}\}$ ,acts on them as $%
Aa_{ij}=(A_{i}+A_{j})a_{ij}$.

For the $AII$ case we chose the field in the form 
\begin{equation}
A=\frac{1}{\sqrt{8}}diag_{2N,2N}\{1,1,-\frac{1}{N-1},...,-\frac{1}{N-1}\}
\label{Efsu}
\end{equation}
(We work with the same normalization for fields and charges, and the meaning
of our normalization choice $\frac{1}{\sqrt{8}}$ will be clear below). Then
the ground state will contain only particles of the type $a_{12}$, since
they have the maximal charge 2. One can see using the explicit form of
projectors (see Appendix) that the scattering process $a_{12}+a_{12}%
\rightarrow a_{12}+a_{12}$ takes place only in the $P_{2\mu _{2}}$ channel
and the S-matrix for it is a prefactor before the parenthesis in (\ref{Susm}%
).

The situation is more complicated in $CII$ case. We chose 
\begin{equation}
A=\frac{1}{\sqrt{8}}diag_{4P,4P}\{1,-1,0,...,0,-1,1,0,...0\}  \label{Efsp}
\end{equation}
with non zero elements on the places $1,2,2P+1,2P+2$. In principle any
combination of particles which is $O(2)$ invariant and has a maximal charge
in the field $A$, can serve as a representative of the ground state. One can
see that the combination $d=a_{1,P+2}-a_{2,P+1}+a_{1,P+1}-a_{2,P+2}$ has the
maximal charge 2. In addition this particle has two important properties,
which one can find analyzing the projectors (see Appendix):firstly, $%
(dd)_{ij}=0$, and hence $P_{\mu _{2}},P_{2\mu _{1}},P_{0}$ are zero for the
scattering of $d$ on itself, and , secondly, this particle scattering on
itself does not produce other particles and amplitude of the scattering is
the coefficient before the projector $P_{2\mu _{2}}$- the prefactor before
the parenthesis in (\ref{Spsm}).

So in both $AII$ and $CII$ cases, with the choice of \ external field we
described above, we have one particle in the ground state. Following
standard technology of thermodynamic Bethe ansatz (TBA), one can get the TBA
equation for the so called dressed energies of the particles 
\begin{equation}
\varepsilon (\theta )=h-m\cosh \theta +\int_{-B}^{B}d\theta ^{\prime }\phi
(\theta -\theta ^{\prime })\varepsilon (\theta ^{\prime })  \label{Inteq}
\end{equation}
in terms of which free energy as a function of external field is 
\begin{equation}
F(h)-F(0)=-\frac{m}{2\pi }\int_{-B}^{B}d\theta \cosh (\theta )\varepsilon
(\theta )  \label{FeTBA}
\end{equation}
Here $B$ is a function of $h/m$ determined by the boundary condition $%
\varepsilon (\pm B)=0$, and $\phi $ is a kernel defined by the S matrix $%
S(\theta )=X(\theta )S_{\min }(\theta )$ for the scattering of the
particles: 
\[
\phi (\theta )=\frac{1}{2\pi i}\frac{d}{d\theta }\ln S(\theta ) 
\]
The Wiener-Hopf method of solution of the integral equation of the form (\ref
{Inteq}) in the limit $h/m\rightarrow \infty $ , aimed to extraction of the
free energy (\ref{FeTBA}), gives an answer for it 
\begin{equation}
F(h)-F(0)=-\frac{\kappa ^{2}}{4}h^{2}\left[ \ln \frac{h}{m}+(s+\frac{1}{2}%
)\ln \ln \frac{h}{m}+c^{\prime }+...\right]  \label{FreeTBA}
\end{equation}
if Fourier transform of the kernel $\widehat{\phi }(\omega )=1-K(\omega )$
factorizes into $K(\omega )=\frac{1}{K_{+}(\omega )K_{-}(\omega )}$, where $%
K_{\pm }(\omega )$ are bounded and have no poles or zeros in the upper(lower)
half plane and have an asymptotic for small $\xi $ 
\begin{equation}
K_{+}(i\xi )=\frac{\kappa }{\sqrt{\xi }}e^{-s\xi \ln \xi }(1-b\xi +O(\xi
^{2}))  \label{Kerasymp}
\end{equation}
with some constants $s$, $\kappa $ and $b$, and 
\begin{equation}
c^{\prime }=\ln \frac{\sqrt{2\pi }\kappa e^{-b}}{K_{+}(i)}-1+s(\gamma
_{E}-1+\ln 8)  \label{Cprime}
\end{equation}
The detailed proof of this statement one can find in \cite{Weisz}.
Calculation of the Fourier transform of the kernels gives 
\begin{equation}
K(\omega )=2e^{\pi |\omega |\Delta }\frac{\sinh (\pi |\omega |\Delta )\sinh
(\pi \omega (1-2\Delta ))}{\sinh (\pi \omega )}  \label{Kersu}
\end{equation}
for $AII$ case and 
\begin{equation}
K(\omega )=2e^{\pi |\omega |\Delta }\frac{\sinh (\pi |\omega |\Delta )\cosh
(\pi \omega (\frac{1}{2}-2\Delta ))}{\cosh (\pi \omega /2)}  \label{Kersp}
\end{equation}
for $CII$ case.
By $|\omega |$ we mean here the function which has the following analytical
continuation to the whole complex plane: $|\omega |$ $\rightarrow \omega
\ast sign(Re(\omega))$.
Factorization of \ (\ref{Kersu}) may be done as 
\[
K_{+}(\omega )=\sqrt{-i\frac{\Delta (1-2\Delta )\omega }{2\pi }}e^{-i\omega
\Delta \ln (-i\omega )+i\mu \omega }\frac{\Gamma (-i\omega \Delta )\Gamma
(-i\omega (1-2\Delta ))}{\Gamma (-i\omega )} 
\]
where the boundness of $K_{+}$ requires $\mu =\Delta \ln \Delta +(1-2\Delta
)\ln (1-2\Delta )$, which leads to the asymptotic behavior of \ the type (%
\ref{Kerasymp}) with the constants 
\[
s=-\Delta ,\ \kappa =\frac{1}{\sqrt{2\pi \Delta (1-2\Delta )}},b=\mu -\Delta
\gamma _{E} 
\]
Kernel (\ref{Kersp}) factorizes as 
\[
K_{+}(\omega )=\sqrt{-i\frac{\Delta \omega }{2\pi }}e^{-i\omega \Delta \ln
(-i\omega )+i\mu \omega }\frac{\Gamma (-i\omega \Delta )\Gamma (\frac{1}{2}%
-i\omega (\frac{1}{2}-2\Delta ))}{\Gamma (\frac{1}{2}-\frac{i\omega }{2})} 
\]
with $\mu =\Delta \ln \Delta +(\frac{1}{2}-2\Delta )\ln (\frac{1}{2}-2\Delta
)+\frac{1}{2}\ln 2$. It has asymptotic (\ref{Kerasymp}) with the following
constants: 
\[
s=-\Delta ,\ \kappa =\frac{1}{\sqrt{2\pi \Delta }},b=\mu +\gamma _{E}\Delta 
\]

\section{Comparison of results}

We are going now to compare the results of perturbative (\ref{FreenL}) and
non perturbative (\ref{FreeTBA}) TBA calculation of the free energy. First
of all, using the form of external fields (and hence the charges) we chose (%
\ref{Efsu}),(\ref{Efsp}), one can see that prefactors of parenthesis
coincide for both $AII$ and $CII$ cases. Although the normalization of
charges is ambiguous total factor, the fact that in this prefactor we get in
both cases the same dependence on $N$ and on $P$, is highly nontrivial check
of not only our S-matrix conjecture, but also of the conjecture about the
particle content of the ground state. Second, even more impressive, check is
the comparison of the coefficients before the subleading term $\ln \ln h$ in
both formulas. One of them is defined by beta function coefficients, another
- by purely exact S-matrix dependent TBA analysis. Again, they coincide for
both $AII$ and $CII$ cases. Moreover, after we saw the coincidence of the
leading and subleading terms in the limits $h>>\Lambda _{MS},h>>m$, one can
use different expressions (\ref{FreenL}) and (\ref{FreeTBA}) for the same
quantity in order to fix the relation between mass scale $m$ and the
renormalization scheme parameter $\Lambda _{MS}$. This comparison involves
the constant terms $c,c^{\prime }$ in both expressions. In the leading order
of big $h$ one just has what is called exact mass gap for the models 
\[
\ln \frac{m}{\Lambda _{MS}}=c-c^{\prime }+O(1/\ln h) 
\]
Using the form of the external field (\ref{Efsu}),(\ref{Efsp}) and the
expressions for $c,c^{\prime }$(\ref{C}),(\ref{Cprime}), one can calculate
the gap explicitly for both sigma models.

\section{\protect\bigskip Discussion}

We proposed fundamental S-matrices for two dimensional sigma models on $AII$
and $CII$ symmetric spaces. We checked them by comparison of perturbative
and TBA calculations of free energy in a strong external field of a specific
form and found the desired correspondence. In this check not all the
particles of the conjectured spectrum have participated - only subsector of
fundamental S-matrix was used. In this sense $T>0$ TBA check based on
extraction of UV central charge seems to be a more complete, since it
involves all the particles of the spectrum.We hope to report on this soon 
\cite{Me}.

The quantum integrability of $AII$ sigma model was expected since the
factorization subgroup is simple in this case, but the integrability of $CII$
sigma model is a ''surprise'' in this sense, because one might expect
anomalies for non local current conservation \cite{AGF}. It is clear that a
deeper understanding of their integrability is desired from the point of
view of conserved currents algebra and their symmetry. More exactly, the
question remaining unclear is what kind of Yangian symmetry is responsible
for integrability of sigma models on coset spaces. For today, the only known
to us mathematically rigorous formulation of coset like symmetric Yangians are
twisted Yangians, but they are known to be responsible for boundary
integrability of sigma models \cite{MacKay}, and hardly have something to do
with quantum integrability without boundary.

\section{Acknowledgments}

The author is grateful to S.Elitzur for many helpful discussions and to
P.Fendley for useful communications.

\section{\protect\bigskip Appendix}

\subsection{Bilinear action}

If we chose the symplectic form as $2N$ by $2N$ matrix of the following
block form $\left( 
\begin{array}{cc}
0 & 1 \\ 
-1 & 0
\end{array}
\right) $, the basis (non orthonormal) for the generators of Lie algebra of $%
AII$ symmetric space one can chose in the form of the following $2N$ by $2N$
matrices 
\begin{eqnarray*}
E_{ij}^{I} &=&(E_{ij}+E_{j+N,i+N}),\ i\neq j \\
E_{ij}^{II} &=&(E_{i,j+N}-E_{j,i+N}) \\
E_{ij}^{III} &=&(E_{i+N,j}-E_{j+N,i}) \\
H_{i} &=&(E_{ii}-E_{i+1,i+1}+E_{i+N,i+N}-E_{i+N+1,i+N+1})
\end{eqnarray*}
where $i,j$ in the first three lines are running from 1 to $N$, and in the
last (Cartan) generators - $i$ runs from 1 to $N-1$. Here $E_{ij}$ is $2N$
by $2N$ matrix with one non zero element equal to 1 and located at the
position $(i,j)$. The motivation for this choice is clear: the first 3 types
of generators with opposite choice of sign between two $E$ belong to $Sp(N)$
since are of the form 
\[
\left( 
\begin{array}{cc}
a & b \\ 
c & -a^{T}
\end{array}
\right) 
\]
required for Lie algebra of $Sp(N)$ with the symplectic form choice we made,
where $a,b,c$-are $N$ by $N$ matrices and $b$ and $c$ - symmetric. So,
opposite choice of signs means that these generators are in orthogonal
completion, i.e. in the coset algebra $su(2N)/sp(N)$. Condition of unitarity
for general $g=$ $\exp \left( i\sum_{I}n^{I}(x)E_{I}\right) $ now reads as $%
n_{ij}^{I\ast }=n_{ji}^{I},n_{ij}^{II\ast }=n_{ji}^{III},n_{ij}^{III\ast
}=n_{ji}^{II}$. It is clear that $n^{II},n^{III}$ can be taken antisymmetric.

Generators of $sp(2P)/sp(P)\times sp(P)$ coset Lie algebra may be chosen in
the form of $4P$ by $4P$ matrices, dividing them into 16 $P$ by $P$ block
matrices.We define the action of the first invariant $sp(P)$ subalgebra in
the four blocks $(1,1),(1,3),(3,1),(3,3)$ then the second $sp(P)$ acts in
the blocks $(2,2),(2,4),(4,2),(4,4)$. We require the generators of the coset
Lie algebra just to be zero in these eight blocks and get the following
basis: 
\begin{eqnarray*}
E_{ij}^{I} &=&E_{i,j+P}-E_{j+3P,i+2P} \\
E_{ij}^{II} &=&E_{i+P,,j}-E_{j+2P,i+3P} \\
E_{ij}^{III} &=&E_{i,j+3P}+E_{j+P,i+2P} \\
E_{ij}^{IV} &=&E_{i+2P,j+P}+E_{j+3P,i}
\end{eqnarray*}
These are non zero in the remaining eight blocks, and are generators of $%
sp(2P)$. The reality condition leads to the requirement $%
n_{I}^{T}=n_{II},n_{III}=n_{IV}$. 

Calculation of bilinear terms may be done by expansion of exponent for coset
group element up to the second order in fields and substitution of it into (%
\ref{Lag2}). For instance for ${\cal L}_{0}$ this leads to ${\cal L}%
_{0}=-Tr(\lambda _{a}\lambda _{b})\partial _{\mu }n_{a}\partial _{\mu }n_{b}$%
, where $\lambda _{a},n_{a}$ - a total set of Lie algebra generators and
corresponding fields. Explicit calculation also shows that the term linear
in $h$ gives total derivatives, and may be omitted. In the same way, terms
containing fields for Cartan generators $n_{i}$ drop out from the term
proportional to $h^{2}$ in (\ref{Lag2}). So fields $n_{i}$ in both cases
decouple and may be omitted. For the remaining fields we change notations and
normalization: $AII$ case $n_{I}\rightarrow n,n_{II}\rightarrow m$, $CII$
case $n_{I}\rightarrow n,n_{III}\rightarrow m$, and get the actions (\ref
{Lag0}).

\subsection{Projectors}

Here we present the projectors appearing in irreducible decomposition of
tensor product of two antisymmetric representations. For projectors
appearing in irrep decomposition of tensor product of two antisymmetric
representations (highest weight $\mu _{2}$) for $SU(2N)$ written in terms
two antisymmetric tensors of rank 2 $a_{ij}$ and $b_{kl}$, one can get using
standard Yang tableau technique. 
\begin{eqnarray}
{\cal P}_{\mu _{4}} &=&\frac{1}{6}%
(a_{ij}b_{kl}+a_{il}b_{jk}+a_{kl}b_{ij}+a_{jk}b_{il}-a_{ik}b_{jl}-a_{jl}b_{ik})
\label{ProjSU} \\
{\cal P}_{2\mu _{2}} &=&\frac{1}{6}%
(2a_{ij}b_{kl}+2a_{kl}b_{ij}+a_{kj}b_{il}+a_{ik}b_{jl}+a_{jl}b_{ik}+a_{il}b_{kj})
\nonumber \\
{\cal P}_{\mu _{1}+\mu _{3}} &=&\frac{1}{2}(a_{ij}b_{kl}-a_{kl}b_{ij}) 
\nonumber
\end{eqnarray}

With more work one can get also projectors appearing in irrep decomposition
of tensor product of two antisymmetric representations (highest weight $\mu
_{2}$) for $Sp(N)$ written in terms two antisymmetric rank 2 tensors $a_{ij}$
and $b_{kl}$ traceless in the sense that $\sum $ $_{i,j}a_{ij}\sigma _{ij}=0$%
, where we chose for symplectic form $\sigma =\left( 
\begin{array}{cc}
0 & 1 \\ 
-1 & 0
\end{array}
\right) $ with four $N\times N$ \ block matrices. 
\begin{eqnarray}
{\cal P}_{\mu _{4}} &=&\frac{1}{6}\left\{
a_{ij}b_{kl}+a_{il}b_{jk}+a_{kl}b_{ij}+a_{jk}b_{il}-a_{ik}b_{jl}-a_{jl}b_{ik}\right.
\label{ProjSp} \\
&&+\frac{4}{N-2}\left( -(ab)_{[ik]}\sigma _{jl}+(ab)_{[il]}\sigma
_{jk}-(ab)_{[jl]}\sigma _{ik}+(ab)_{[jk]}\sigma _{il}\right.  \nonumber \\
&&\left. \left. +(ab)_{[ij]}\sigma _{kl}+(ab)_{[kl]}\sigma _{ij}\right) +%
\frac{4((ab))}{(N-1)(N-2)}(\sigma _{ik}\sigma _{jl}-\sigma _{il}\sigma
_{jk}-\sigma _{ij}\sigma _{kl})\right\}  \nonumber \\
{\cal P}_{\mu _{1}+\mu _{3}} &=&\frac{1}{2}\left[ a_{ij}b_{kl}-a_{kl}b_{ij}+\frac{4%
}{N-2}\left( (ab)_{(il)}\sigma _{jk}+(ab)_{(jk)}\sigma _{il}\right. \right. 
\nonumber \\
&&\left. \left. -(ab)_{(ik)}\sigma _{jl}-(ab)_{(jl)}\sigma _{ik}\right) 
\right]  \nonumber \\
{\cal P}_{2\mu _{2}} &=&\frac{1}{6}\left\{
2a_{ij}b_{kl}+2a_{kl}b_{ij}+a_{kj}b_{il}+a_{ik}b_{jl}+a_{jl}b_{ik}+a_{il}b_{kj}\right.
\nonumber \\
&&+\frac{1}{N-2}\left[ (4(ab)_{ki}+2(ab)_{ik})\sigma
_{jl}-(4(ab)_{li}+2(ab)_{il})\sigma _{jk}\right.  \nonumber \\
&&\left. +(4(ab)_{lj}+2(ab)_{jl})\sigma _{ik}-(4(ab)_{kj}+2(ab)_{jk})\sigma
_{il}-4(ab)_{[kl]}\sigma _{ij}\right]  \nonumber \\
&&\left. +\frac{((ab))}{(N-1)(N-2)}(2\sigma _{ik}\sigma _{jl}-2\sigma
_{il}\sigma _{jk}+4\sigma _{ij}\sigma _{kl})\right\}  \nonumber \\
{\cal P}_{2\mu _{1}} &=&B_{ik}\sigma _{jl}-B_{il}\sigma _{jk}+B_{jl}\sigma
_{ik}-B_{jk}\sigma _{il}  \nonumber \\
{\cal P}_{\mu _{2}} &=&C_{ik}\sigma _{jl}-C_{il}\sigma _{jk}+C_{jl}\sigma
_{ik}-C_{jk}\sigma _{il}  \nonumber \\
{\cal P}_{0} &=&\frac{((ab))}{N(N-1)}(\sigma _{ik}\sigma _{jl}-\sigma _{jk}\sigma
_{il})  \nonumber
\end{eqnarray}

where\bigskip 
\begin{eqnarray*}
(ab)_{ij} &=&\sum_{k,l=1}^{N}a_{ik}b_{jl}\sigma _{kl},\
((ab))=\sum_{i,j=1}^{N}(ab)_{ij}\sigma _{ij}, \\
B_{ij} &=&\frac{1}{2(N-2)}\left( (ab)_{ij}+(ab)_{ji}\right) ,\ C_{ij}=\frac{1%
}{2(N-2)}\left( (ab)_{ij}-(ab)_{ji}-\frac{2}{N}((ab))\right)
\end{eqnarray*}

\end{document}